\documentclass[12pt]{article}
\usepackage[psamsfonts]{amsfonts,amssymb}
\usepackage[psamsfonts,mathscr]{eucal}

\newcommand{\Def}{\newcommand}
\Def{\DeF}{\renewcommand}
\Def{\Thm}{\newtheorem}
\Def{\bq}{\begin{equation}}
\Def{\eq}{\end{equation}}
\Def{\bQ}{\begin{eqnarray}}
\Def{\eQ}{\end{eqnarray}}
\Thm{definition}{Definition}
\Thm{application}{Application}
\Thm{hypothesis}{Hypothesis}
\Thm{lemma}{Lemma}
\Thm{problem}{Problem}
\Thm{proposition}{Proposition}
\Thm{remark}{Remark}
\Thm{theorem}{Theorem}
\DeF{\AA}{\mathscr{A}}
\Def{\BB}{\mathscr{B}}
\Def{\CC}{\mathscr{C}}
\Def{\DD}{\mathscr{D}}
\Def{\GG}{\mathcal{G}}
\Def{\HH}{\mathscr{H}}
\Def{\LL}{\mathscr{L}}
\Def{\NN}{\mathscr{N}}
\Def{\OO}{\mathscr{O}}
\Def{\RR}{\mathscr{R}}
\DeF{\SS}{\mathscr{S}}
\Def{\TT}{\mathscr{T}}
\Def{\bbC}{\mathbb{C}}
\Def{\bbN}{\mathbb{N}}
\Def{\bbR}{\mathbb{R}}
\Def{\bbT}{\mathbb{T}}
\Def{\bbZ}{\mathbb{Z}}
\Def{\And}{\mathrm{and}}
\Def{\Der}{\mathrm{Der}}
\Def{\ID}{\mathbf{1}}
\Def{\ie}{\mathrm{i.e.\,}}
\Def{\If}{\mathrm{If}}
\Def{\ii}{\mathbf{i}}
\Def{\Ker}{\mathrm{Ker\,}}
\Def{\non}{\nonumber}
\Def{\QED}{\emph{Q.E.D.} \mbox{\rule[-1.5pt]{6pt}{10pt}}}
\Def{\Rg}{\mathrm{Rg\,}}
\Def{\st}{\mathrm{s.t.\,}}
\Def{\Then}{\mathrm{Then}}
\Def{\where}{\mathrm{where}}
\Def{\with}{\mathrm{with}}
\DeF{\theequation}{\thesection.\arabic{equation}}
\DeF\refname{}
\title{\rule{0cm}{2cm} \textbf{
Perturbation Theory and Control\\
in Classical or Quantum Mechanics\\
by an Inversion Formula
}}
\author{
Michel VITTOT\\
\vspace{-0.5cm}\\
Centre de Physique Th\'eorique {\small (UMR 6207 \& FRUMAM)}\\
CNRS Luminy, case 907 -- F-13288 Marseille cedex 9 -- France\\
{\small vittot@cpt.univ-mrs.fr\ \ \ \ http://www.cpt.univ-mrs.fr}
}
\date{\today}
\begin{document}
\maketitle
\thispagestyle{empty}
\begin{abstract}
We consider a perturbation of an ``integrable'' Hamiltonian and give
an expression for the canonical or unitary transformation which
``simplifies'' this perturbed system. The problem is to invert a
functional defined on the Lie-algebra of observables. We give a bound
for the perturbation in order to solve this inversion. And apply this
result to a particular case of the control theory, as a first example,
and to the ``quantum adiabatic transformation'', as another example.
\end{abstract}

\newpage
\tableofcontents
\vspace{0.2cm}
\section{- Algebraic Framework}
\label{S01}
\setcounter{equation}{0}
We start with a vector space \( \AA \) which we call the space of
observables.

We will apply this theory to 2 main examples: the classical mechanics,
in which \( \AA \) is an (abelian) algebra of functions defined on the
phase space, and the quantum mechanics, in which \( \AA \) is an
algebra of operators on some Hilbert space. By taking a basis of
projectors in \( \AA \) we will consider these operators as a
collection of ``matrix elements'', i.e. also as functions.

We assume that \( \AA \) is endowed with a Lie stucture, i.e. there
exists a mapping (called the ``bracket'') from \( \AA \) into the
space \( \LL(\AA) \) of linear operators of \( \AA \):
\bQ
& \{\ldots\}: & \AA \to \LL(\AA) \non\\
&& {V} \mapsto \{{V}\} \label{001}
\eQ
which satisfies:
\bQ
\forall {V}, {W} \in \AA\ \ \ \ \{{V}\}{W} &=& - \{{W}\}{V}
\label{002}
\\ \And\ \ \ \{\{{V}\}{W}\} &=& \{{V}\} \{{W}\} - \{{W}\}
\{{V}\} \label{003}
\eQ
so that the ``bracket'' mapping is linear in its argument. It is also
antisymmetric, by (\ref{002}) and satisfies the Jacobi identity
(\ref{003}). We consider a fixed element \( {H} \) of \( \AA \) which
we call a Hamiltonian. The ``motion'' or the ``flow'' generated by \(
{H} \) is the 1-parameter group:
\bq
\forall {t} \in \bbR\ \ \ \ e^{{t} \{{H}\}}: \AA \to \AA \label{004}
\eq
The exponential of \(\ \{{H}\}\ \) is defined by the usual power
series:
\bq
e^{{t} \{{H}\}}\ := \sum_{n = 0}^{\infty} {\{{{t}. H}\}^{n} \over
{n}!} \label{081}
\eq
and is an automorphism of the Lie structure, as is proven below in
(\ref{033}).

Actually what is only needed is that \( \{{H}\} \) is a linear
operator ``affiliated'' to \( \AA \) which means that (\ref{004}) is
still valid: \( \{{H}\} \) is assumed to be the generator of a
1-parameter group of automorphisms, even if \( {H} \) is not an
element of \( \AA \).

\vspace{0.2cm}
Our problem is to find a relation (for instance a ``conjugation'')
between the flow generated by a perturbation of \( {H} \), denoted \(
{H} + {V} \) for some \( {V} \in \AA \), and the flow of \( {H} \).
Poincar\'e called this problem ``the main problem of the dynamics''.
Of course this relation would be mainly useful if we have some
information on the flow of the unperturbed hamiltonian \( {H} \).

\vspace{0.2cm}
We will study 2 types of problems:

\begin{problem} -
If we are permitted to modify the perturbed hamiltonian \( {H} + {V}
\) by adding a ``small'' term \( {f}({V}) \), we will try to find a
relation between the flow of \(\ {H} + {V} + {f}({V}) \) and the flow
of \( {H} \). Then the term \( {f}({V}) \) will be called a
``control'' term.
\end{problem}
Of course we want to exclude the trivial solution \( {f}({V}) = -{V}
\) so we have added the supplementary condition on this term to be
``small'', for instance quadratic in \( {V} \). So we stabilize the
perturbed hamiltonian by a small control term, well adapted to the
problem.

\begin{problem} -
If we are not permitted to modify the perturbed hamiltonian (for
instance if we investigate the motion of some planet), or we don't
want to modify it, then we will try to find a good ``change of
coordinates'', i.e. an automorphism of \( \AA \), which connects the 2
flows: perturbed and unperturbed.
\end{problem}
We will search this automorphism under the form of some exponential \(
e^{\{\Gamma {W}\}}\ \) where we define \(\ \Gamma\ \) as follows.

\vspace{0.2cm}
Let us make an important assumption on \( {H} \), which will be
satisfied when \( {H} \) is ``integrable'', as seen below:
\begin{hypothesis} -
We assume that there exists a linear operator \(\ \Gamma: \AA \to \AA\ \)
such that:
\bq
\{{H}\}^{2}\ \Gamma = \{{H}\}
\label{121}
\eq
\end{hypothesis}
and then we build two other operators \(\ \NN\ \&\ \RR\ \) by:
\bq
\NN := \{{H}\} \Gamma\ \ \ \ \ \RR := \ID - \NN
\label{096}
\eq
Hence \( \Gamma \) is a pseudo-inverse of \( \{{H}\} \):
Let us remind that it is impossible to find a strict inverse of \(
\{{H}\} \) since it always has a non-trivial kernel (for instance \(
\{ {H} \} {H} = 0 \)).

Any element \( {V} \) of \( \AA \) such that \( \{{H}\}{V} =
0 \), is constant under the flow of \( {H} \):
\bq
e^{{t} \{{H}\}} {V} = {V}
\label{037}
\eq
So the vector space \( \Ker\{{H}\} \) is called the set of ``constants
of motion''. It is a sub-Lie-algebra of \( \AA \) since (using
(\ref{003})):
\bq
\If\ \ \ \{{H}\}{V} = \{{H}\}{W} = 0\ \ \ \Then\ \ \ \{{H}\}\{{V}\}{W}
= \{{V}\}\{{H}\}{W} + \{\{{H}\}{V}\}{W} = 0 + 0 \label{006}
\eq
Let us also note that (\ref{121}) can be rewritten as:
\bq
\{{H}\} \RR = 0
\label{103}
\eq
which means that the range \( \Rg \RR \) of the operator \( \RR \) is
included in \( \Ker\{{H}\} \). The \( \ID \) is the identity in the
algebra \( \LL(\AA) \) of endomorphisms of \( \AA \). The notation \(
\RR \) designates the ``resonant part'' and the notation \( \NN \) the
``non-resonant part''. Let us remind that \(\ \{{H}\}, \RR, \NN, \Gamma\
\) are elements of \(\ \LL(\AA) \).

\vspace{0.2cm}
\begin{application} - Classical Mechanics:
\end{application}
\( \AA \) is the algebra of \( \CC^{\infty} \) real-valued functions
of \( (p,q) \) in some domain of \( \bbR^{2{L}} \) or functions of \(
({A},\theta) \in \HH \times \bbT^{L} \) for some domain \( \HH \) of
\( \bbR^{L} \) ({L} is the number of degree of freedoms). It is more
convenient to make a Fourier transformation in \( \theta \), and so \(
\AA \) may be taken as the algebra of functions \( {V} \) of \(
({A},\Delta) \in \HH \times \bbZ^{L} \) into \( \bbC \), such that \(
{V}({A},-\Delta) = {V}({A},\Delta)^{*} \). The Lie structure is the
Poisson bracket. We can compute the operators \( \RR \) and \( \Gamma
\) when \( {H} \) is ``integrable'', i.e. when \( {H} \) is a function
of the ``actions variables'' only, and does not depend on the ``angles
variables'':
\bq
{H}({A},\theta) = {h}({A}) \label{022}
\eq
or after a Fourier transformation in \( \theta \):
\bq
{H}({A},\Delta) = {h}({A}).\delta_{0}(\Delta) \label{023}
\eq
with \( \delta \) the Kronecker symbol. Indeed, let us denote by \(
\omega({A}) \) the derivative (gradient) of \( {h} \) with respect to
\( {A} \). Then the bracket \( \{{H}\} = \omega({A}). \partial_{
\theta}\ \) and so, after a Fourier transformation, it is given by:
\bq
(\{{H}\}{V})({A},\Delta) = \ii. (\omega({A}) \cdot \Delta).
{V}({A},\Delta) \label{040}
\eq
on any element \( {V} \) of \( \AA \). We have used a scalar product
\( \omega({A}) \cdot \Delta\ \). So that the operator \( \RR \) is:
\bq
(\RR {V})({A}, \Delta) = {V}({A}, \Delta). \chi( \omega({A}) \cdot
\Delta = 0) \label{041}
\eq
where we introduce a characteristic function \(\ \chi(\omega({A})
\cdot \Delta = 0)\ \) which is 1 when \(\ \omega({A}) \cdot \Delta =
0\ \) and is 0 otherwise. We could have written it as \(\ \chi( \Delta
\in \omega({A})^{ \perp} \cap \bbZ^{L}) \). Similarly:
\bq
(\NN {V})({A},\Delta) = {V}({A},\Delta). \chi (\omega( {A}) \cdot
\Delta \neq 0) \label{039}
\eq
Then the action of the operator \( \Gamma \) is given by:
\bq
(\Gamma {V})({A},\Delta):= {\chi (\omega({A}) \cdot \Delta \neq 0)
\over \ii.(\omega({A}) \cdot \Delta)} \cdot {V} ({A}, \Delta)
\label{021}
\eq
See section \ref{S03} for more details and section \ref{S04} where we
introduce a norm.

\vspace{0.2cm}
\begin{application} - Quantum Mechanics:
\end{application}
\( \AA \) is the algebra of operators on some separable Hilbert space.
More precisely we assume that \( \AA \) has a unity (denoted by \( \ID
\)), and we consider a maximal family of mutually orthogonal
projectors \( {P}_{A} \in \AA\ \) where \(\ {A}\ \) varies in some
countable set \( \HH \). That means:
\bQ
\sum_{{A} \in \HH} {P}_{A} &=& \ID \label{012} \\
\forall {A}, {{A}'} \in \HH\ \ \ {P}_{A}.{P}_{{A}'} &=& {P}_{A}.
\delta_{{A}, {{A}'}} \label{011}
\eQ
with \( \delta \) the Kronecker symbol. \( \RR \) and \( \Gamma \) are
easily written when \( {H} \) can be ``diagonalized''. We choose \(
{H} \) as follows:
\bq
{H} = \sum_{{A} \in \HH} {h}({A}).{P}_{A} \label{013}
\eq
for some function \( {h}: \HH \to \bbR \). So \( {h}(\HH) \) is the
spectrum of \( {H} \) and the projectors are the spectral projectors
of \( {H} \). We can choose \(\ {h}\ \) to be the identity function by
taking \(\ \HH \ \) as the spectrum of \(\ {H} \).

\begin{remark} -
Reciprocally, if \( \AA \) were endowed with an involution (``*'',
named ``adjonction'') we could have started by giving a self-adjoint
operator \( {H} \) with pure point spectrum, and then we could have
taken the spectral projectors of \( {H} \), as the family of
projectors \( {P}_{A} \).
\end{remark}

Then any element \( {V} \) of \( \AA \) can be written as:
\bq
{V} = \sum_{{A}, {{A}'} \in \HH} {V}_{{A},{{A}'}}\ \ \ \where\ \ \
{V}_{{A},{{A}'}}:= {P}_{A} {V} {P}_{{A}'} \label{015}
\eq
It is convenient to introduce the set:
\bq
\GG({A}):= \HH - {A}:= \{{{A}'} - {A}\}_{{{A}'} \in \HH} \label{045}
\eq
and to define:
\bq
\forall {A} \in \HH\ \ \ \forall \Delta \in \GG({A})\ \ \
{V}({A},\Delta):= {P}_{{A} + \Delta} {V} {P}_{A} \label{043}
\eq
i.e. \( {V}({A},\Delta) = {V}_{{A} + \Delta,{A}} \). For instance:
\bq
{H}({A},\Delta) = {h}({A}).\delta_{0}(\Delta).{P}_{A} \label{100}
\eq
which is similar to (\ref{023}).

The Lie structure is given by the commutator. Actually we multiply it
by \( \ii/\hbar \) for some constant \( \hbar \) which has the correct
dimensionality, so that \(\ {t} \{{H}\} \) has no dimension:
\bq
\forall {V}, {W} \in \AA\ \ \ \{{V}\}({W}):= \ii ({V}.{W} - {W}.
{V})/\hbar \label{018}
\eq
Let us now turn to the action of the operators \( \{{H}\} \) and \(
\RR \):
\bq
(\{{H}\}{V})({A},\Delta) = \ii. \Bigl( {h}({A} + \Delta) - {h}({A})
\Bigr). {V}({A},\Delta)/\hbar \label{048}
\eq
Hence:
\bq
(\RR {V})({A},\Delta) = {V}({A},\Delta).\ \chi \Bigl( {h}({A} +
\Delta) = {h}({A}) \Bigr) \label{050}
\eq
and:
\bq
(\NN {V})({A},\Delta) = {V}({A},\Delta).\ \chi \Bigl( {h}({A} +
\Delta) \neq {h}({A}) \Bigr) \label{049}
\eq
Then the operator \( \Gamma \) is given by:
\bq
(\Gamma {V})({A},\Delta) = {\hbar.\chi \Bigl({h}({A} + \Delta) \neq
{h}({A}) \Bigr) \over \ii. \Bigl( {h}({A} + \Delta) - {h}({A}) \Bigr)}
\cdot {V}({A},\Delta) \label{019}
\eq
See section \ref{S03} for more details and section \ref{S04} where we
introduce a norm.

\vspace{0.2cm}
Remark: we have introduced an arbitrary constant \(\ \hbar\ \) in
(\ref{018}). But we will notice that the operator \( \Gamma \) always
appears with a bracket around it. And the pair \( \{ \Gamma \ldots \}
\) is independant of \(\ \hbar\ \), since this constant is in the
numerator and the denominator.

\section{- The Main Theorems}
\label{S08}
\setcounter{equation}{0}
\hrule\mbox{ }
\begin{theorem} -
The control problem (problem 1) is solved by an explicit formula. Let
us first define the functions \( {F}\ \And\ {f} : \AA \to \AA \) by:
\bQ
{F}({V}) &:=& e^{-\{\Gamma {V}\}} \RR {V} + {1 - e^{-\{\Gamma {V}\}}
\over \{\Gamma {V}\}} \NN {V} \label{008}\\
{f}({V}) &:=& {F}({V}) - {V} \label{009}
\eQ
Then we have:
\bq
\forall {t} \in \bbR\ \ \ \ \ e^{{t} \{{H} + {V} + {f}({V})\}} =
e^{-\{\Gamma {V}\}}. e^{{t} \{{H}\}}. e^{{t} \{\RR {V}\}}. e^{\{\Gamma
{V}\}} \label{010}
\eq
\hrule
\end{theorem}
The meaning of the second term in (\ref{008}) is the following:
\bq
{1 - e^{-\{\Gamma {V}\}} \over \{\Gamma {V}\}} = \sum_{n \in \bbN}
{\{-\Gamma {V}\}^{n} \over n + 1!} = \int_{0}^{1}ds. e^{- s \{\Gamma
{V}\}} \label{024}
\eq
We will prove this theorem in section \ref{S02}. This will solve the
control problem if we can check that \( {f}({V}) \) is indeed smaller
than \( {V} \). This will be proved in section \ref{S04}. To see this,
it is sufficient to notice that in the expression (\ref{008}) of \(
{F}({W}) \), the terms \( e^{-\{\Gamma {W}\}} \) and \( (1 -
e^{-\{\Gamma {W}\}}) / \{\Gamma {W}\}\ \) are near \( \ID \) when \(
{W} \approx 0\). So:
\bq
{F}({W}) \approx \RR {W} + \NN {W}:= {W}\ \ \ \mathrm{so\ that}\ \ \
{f}({W}) = \OO({W}^{2}) \label{027}
\eq
as is seen in (\ref{059}).

The formula (\ref{010}) connects the perturbed flow, modified by a
control term, with the unperturbed flow.

The new factor, the flow of \( \RR {V} \) will turn out to commute
with the flow of \( {H} \): cf. (\ref{036}).

\vspace{0.2cm}
The second problem (the ``change of coordinates'') is solved by an
inversion formula. Let us rewrite (\ref{010}) with \( {F}({V}) \)
instead of \( {V} + {f}({V}) \), and with \( {W} \) instead of \( {V}
\):
\bq
\forall {t} \in \bbR\ \ \ \ \ e^{{t} \{{H} + {F}({W})\}} =
e^{-\{\Gamma {W}\}}.e^{t \{{H}\}}. e^{{t} \{\RR {W}\}}.e^{\{\Gamma
{W}\}} \label{025}
\eq

\newpage
\hrule
\begin{theorem} -
If we can find \( {W} \) such that \( {F}({W}) = {V},\ \ie\ {W} =
{F}^{-1}({V}) \), then:
\bq
\forall {t} \in \bbR\ \ \ \ \ e^{{t} \{{H} + {V}\}} = e^{-\{\Gamma
{W}\}}.e^{t \{{H}\}}. e^{{t} \{\RR {W}\}}.e^{\{\Gamma {W}\}}
\label{014}
\eq
\hrule
\end{theorem}
Here again, the flow of \( \RR {W} \) commute with the flow of \( {H}
\).

Hence we need to invert the function \( {F} \). But \( {F} \) is near
the identity function around 0. More precisely we will find a ball
around 0 in \( \AA \), for some norm, such that the difference between
\( {F} \) and the identity function (what we called \( {f} \) in
(\ref{009})) is Lipschitz and contractant: cf. (\ref{027}). So \( {F}
\) can be inverted, at least around 0. See section \ref{S04}.

\vspace{0.2cm}
So to summarize this introduction: the first problem is solved
explicitly for any perturbation, but we need to assume some smallness
on the size of the perturbation to ensure that the ``control'' term is
smaller than the original perturbation. And the second problem is
solved by an inversion formula, which also needs some smallness on the
size of the perturbation to ensure that \( {F} \) is invertible. To be
more precise, we need some norm on the Lie-algebra \( \AA \): this is done
in section \ref{S04}.

\vspace{0.2cm}
The equation in the unknown \( {W} \) (i.e. the inversion \( {W} =
{F}^{-1}({V}) \)) may be named the ``Hamilton-Jacobi equation''.
Indeed it yields the automorphism \( e^{\{\Gamma {W} \}}
\).

\vspace{0.2cm}
\begin{remark}
If we replace the hypothesis 1 by a (slightly) stronger one,
then the situation is clearer, and simpler:

\vspace{0.2cm}
We assume that there exists a linear operator \(\ \Gamma: \AA \to \AA\ \)
such that:
\bQ
&& \{{H}\}^{2}\ \Gamma = \{{H}\} \Gamma \{{H}\} = \{{H}\} \label{182}\\
&& \{{H}\}\ \Gamma^{2} = \Gamma \{{H}\} \Gamma = \Gamma
\label{178}
\eQ
And we define 4 operators:
\bQ
&& \NN := \{{H}\} \Gamma\ \ \ \ \ \RR := \ID - \NN \\
&& \tilde{\NN} := \Gamma \{{H}\}\ \ \ \ \ \tilde{\RR} := \ID -
\tilde{\NN} \label{179}
\eQ
Under this stronger assumption, we easily show that:
\bQ
&& \NN^{2} = \NN\ \ \ \ \ \RR^{2} = \RR\ \ \ \ \ \tilde{\NN}^{2} =
\tilde{\NN}\ \ \ \ \tilde{\RR}^{2} = \tilde{\RR}\\
&& \Ker \tilde{\RR} = \Rg \tilde{\NN} = \Rg \Gamma \subset \Ker \RR =
\Rg \NN = \Rg \{{H}\}\\
&& \Ker \NN = \Rg \RR = \Ker \Gamma \subset \Ker \tilde{\NN} = \Rg
\tilde{\RR} = \Ker \{{H}\}
\label{180}
\eQ
so that we have a characterization of the sub-Lie-Algebra \(\ \Ker
\{{H}\} \): the ``constants of the motion'' are exactly \( \Rg
\tilde{\RR} \).

Furthermore, the 2 above ``set-inequalities'' become
``set-equalities'' if and only if:
\bq
\{ {H} \} \Gamma = \Gamma \{ {H} \}\ \ \ \ \ie\ \ \ \ \tilde{\NN} = \NN
\label{181}
\eq
In that case (\ref{182}, \ref{178}) are equivalent to (\ref{121}). For
instance, the assumption (\ref{181}) is satisfied in examples
(\ref{021}) or (\ref{019}).

But we don't need this stronger hypothesis for the rest of this paper.
\end{remark}

\section{- Localization of the Action Variable}
\label{S10}
\setcounter{equation}{0}
Let us study the case of the classical mechanics, as above, with
action-angles variables. Usually, we localize the action variable \(
{A} \) near some point \( {A}_{0} \) i.e. we make the canonical change
of variables, from \( ({A}, \theta) \) to \( ({A}_{1}, \theta_{1}) \)
with:
\bq
{A} = {A}_{0} + \varepsilon {A}_{1}\ \ \ \ \ \ \theta = \theta_{1}
\label{153}
\eq
along with a rescaling of the hamiltonian:
\bq
{H}_{1}({A}) := {H}({A}_{0} + \varepsilon.{A})/ \varepsilon
\label{154}
\eq
where we choose some positive constant \( \varepsilon \). This
transformation is called a ``canonical similarity'' since it preserves
the symplectic form, up to the multiplicative constant \( \varepsilon
\). When \( {H} \) is integrable in the usual sense, with
actions-angles variables, we can expand the hamiltonian \( {H} =
{H}({A}) \) around \( {A}_{0} \):
\bq
{H}({A}_{0} + \varepsilon {A}_{1}) = {c} + \varepsilon. \omega \cdot
{A}_{1} + {q}(\varepsilon. {A}_{1})\ \ \ \where\ \ \ \omega :=
{H}'({A}_{0})\ \ \ \And\ \ \ {q}(0) = {q}'(0) = 0 \label{155}
\eq
(i.e. \(\ {q}\ \) is quadratic in \( {A}_{1} \)) and where the
additive constant \( {c} := {H}({A}_{0}) \) is not relevant and will
be forgotten. Hence:
\bq
{H}_{1}({A}) = \omega \cdot {A} + \varepsilon. \tilde{q}({A})
\label{156}
\eq
where \( \tilde{q}({A}) := {q}(\varepsilon. {A})/\varepsilon^{2}\ \)
is of order \( \varepsilon^{0} \).

Let us introduce a perturbation \( {V} \), as above, and choose \(
\varepsilon := ||{V}||^{1/2} \) for some norm. Under the rescaling
(\ref{154}) of the hamiltonian, the perturbation is also divided by \(
\varepsilon \) and so becomes:
\bq
{V}_{1}({A}, \theta) := {V}({A}_{0} + \varepsilon.{A}, \theta)/
\varepsilon \label{157}
\eq
which is of order \( \varepsilon \), so that the perturbed hamiltonian
is now:
\bq
{H}_{1}({A}) + {V}_{1}({A}, \theta) = \omega \cdot {A} + \varepsilon.
{V}_{2}({A}, \theta) \label{158}
\eq
\bq
\with\ \ \ \ {V}_{2}({A}, \theta) = \tilde{q}({A}) + {V}_{1}({A},
\theta)/ \varepsilon \label{159}
\eq
which is of order \( \varepsilon^{0} \). So to summarize, it is always
possible to assume that the integrable part is an harmonic oscillator,
at least locally in the variable \( {A} \), i.e. in a region (around
any fixed \( {A}_{0} \)) where \( {A} \) (in the hamiltonian
(\ref{158})) is of order \( \varepsilon^{0} \), but it may be wrong
when \( {A} \) is of order \( 1/\varepsilon \). Hence in this case of
classical mechanics, localized in action variable, the operator \(\
\Gamma\ \) is:
\bq
\Gamma = {1 \over \omega \cdot \partial} \cdot \NN\ \ \ \with\ \ \
\partial := \partial_{\theta} \label{162}
\eq
Let us explicit the action of \(\ \Gamma \) on an arbitrary
trigonometric observable:
\bq
\Gamma e^{\ii \theta \Delta} = {e^{\ii \theta \Delta} \over \ii
\omega \Delta} \cdot \chi (\Delta \in \bbZ^{L} \setminus \omega^{\perp})
\label{005}
\eq
So that the operator \( \NN \) given by (\ref{039}) becomes here,
after a Fourier transformation from the angles \( \theta \) to the
integer vector \( \Delta \):
\bq
(\NN {V})({A},\Delta) = {V}({A},\Delta). \chi (\Delta \in \bbZ^{L}
\setminus \omega^{\perp}) \label{163}
\eq

\section{- Non-Resonant Hamiltonians}
\label{S00}
\setcounter{equation}{0}
We can also define the notion of ``non-resonance'' as follows.
\vspace{0.2cm}

\hrule
\begin{definition} -
In classical mechanics, {H} is ``non-resonant'' iff:
\bq
\forall\ {V}, {W}\ \in \Ker\{H\}\ \ \ \mathrm{we\ have}\ \ \{{V}\} {W}
= 0\label{160}
\eq
In quantum mechanics, {H} is ``non-resonant'' iff:
\bq
\forall\ {V}, {P}\ \in \Ker\{H\}\ \ \st\ {P}^{2} = {P}\ \ \ \mathrm{we\
have}\ \ \{{V}\} {P} = 0 \label{161}
\eq
\end{definition}
\hrule\mbox{ }

Hence in the classical case:
\bq
\forall {W}, {V} \in \AA\ \ \ \{\RR {W}\}\RR {V} = 0\ \ \ \ \ie\ \ \
\{\RR \AA\}\RR = 0 \label{082}
\eq
The above definition means that any 2 constants of motion of \( {H} \)
do commute together.

After the localisation of the action variable, as in the preceding
section \ref{S10}, the above non-resonant condition (\ref{160}) is
equivalent to the ``usual'' non-resonance condition, which is in
classical mechanics:
\bq
\omega^{\perp} \cap \bbZ^{L} = \{ 0 \} \label{129}
\eq
i.e. there are no integer vector \( \Delta \) orthogonal to \( \omega
\), except for \( \Delta = 0 \). And the operator \( \RR \) may be
written as the multiplication by the characteristic function \(
\chi(\Delta = 0) \).

We will give an example of a resonant hamiltonian in (\ref{074}), for
which we can still apply our control theory.

\vspace{0.2cm}
In quantum mechanics, the non-resonance condition (\ref{161}) means
that any spectral projector of \( {H} \) commutes with any constant of
motion. The non-resonance condition is ``usually'' defined for the
Floquet case:
\bq
{H} = \sum_{{k} \in \bbZ, {A} \in \bbN} ({k} + {h}({A})). {P}_{{k},
{A}} \label{130}
\eq
for some projectors \( {P}_{{k}, {A}} \) which are mutually
orthogonal. For instance the function \( {h}({A}) \) may be taken as
\(\ \omega. {A}\ \) or \(\ \omega. {A}^{2} \). The condition
(\ref{161}) is satisfied exactly when the set of the spectral gaps
intersects \( \bbZ \) in the only point 0:
\bq
\GG \cap \bbZ = \{ 0 \}\ \ \ \where\ \ \ \GG := \{ {h}({A}) -
{h}({B})\ \ \st\ \ {A}, {B} \in \bbN \} \label{164}
\eq
For the case where \( {h}({A}) = \omega. {A}^{a}\ \) for some positive
integer \( {a} \), the condition (\ref{164}) exactly means that the
``frequency'' \( \omega \) is ``non-resonant'' in the usual sense.
We see that the ``resonance'' condition is a different notion than the
degeneracy property, i.e. the dimension of the spectral projectors.

\vspace{0.2cm}
Let us note that the r.h.s. of (\ref{014}) is called the ``normal
form'' of the (perturbed) flow on the l.h.s. When \( {H} \) is
``resonant'', then (\ref{014}) is called the ``resonant normal form''.

\section{- Proof of Theorems 1 \& 2}
\label{S02}
\setcounter{equation}{0}
Actually the theorems (\ref{010}) \& (\ref{014}) are 2 different
interpretations of the same formula (\ref{010}) or (\ref{025}). Let us
first prove:

\vspace{0.6cm}
\hrule
\begin{proposition}
\bq
\forall {W} \in \AA\ \ \ \ {H} + {F}({W}) = e^{-\{\Gamma {W}\}} ({H} +
\RR {W}) \label{028}
\eq
\hrule
\end{proposition}

\textbf{Proof}: indeed from the definition (\ref{008}) and
(\ref{096}), \( {F}({W}) \) can be rewritten as:
\bq
{F}({W}) - e^{-\{\Gamma {W}\}} \RR {W} = {1 - e^{-\{\Gamma {W}\}}
\over \{\Gamma {W}\}} \{{H}\} \Gamma {W} = - {1 - e^{-\{\Gamma {W}\}}
\over \{\Gamma {W}\}} \{\Gamma {W}\} {H} = e^{-\{\Gamma {W}\}} {H} -
{H} \label{029}
\eq
where we used the antisymmetry (\ref{002}). \QED

\vspace{0.2cm}
\textbf{Proof of (\ref{025})}: Let us now take the brackets of the 2
sides of (\ref{028}):
\bq
\{{H} + {F}({W})\} = \{e^{-\{\Gamma {W}\}} ({H} + \RR
{W})\}\label{030}
\eq
But:
\bq
\forall {V},{W} \in \AA\ \ \ \ \{e^{\{{V}\}}{W}\} = e^{\{{V}\}}.
\{{W}\}. e^{-\{{V}\}} \label{033}
\eq
Indeed:
\bq
\forall {V},{W} \in \AA,\ \ \forall n \in \bbN\ \ \ \
\{\{{V}\}^{n}{W}\} = \sum_{k = 0}^{n} {n \choose k}. \{{V}\}^{n - k}.
\{{W}\}. \{-{V}\}^{k} \label{035}
\eq
The proof of (\ref{035}) is an easy recurrence from the case \( n = 1
\): cf (\ref{003}). Hence:
\bq
\{{H} + {F}({W})\} = e^{-\{\Gamma {W}\}}.\{{H} + \RR {W}\}.e^{\{\Gamma
{W}\}} \label{034}
\eq
Let us now exponentiate the 2 sides of (\ref{034}) (multiplied by any
\( {t} \in \bbR \)):
\bq
e^{t\{{H} + {F}({W})\}} = \exp \Bigl[ {{t}. e^{-\{\Gamma {W}\}}. \{{H}
+ \RR {W}\}. e^{\{\Gamma {W}\}}} \Bigr] = e^{- \{\Gamma {W}\}}. e^{{t}
\{{H} + \RR {W}\}}. e^{\{\Gamma {W}\}} \label{031}
\eq
where we have used:
\bq
\forall A, B\ \ \ \ e^{A^{-1}.B.A} = A^{-1}.e^{B}.A \label{032}
\eq
To finish the proof of (\ref{025}), there remains to show that \(
\{{H}\} \) and \( \{{\RR {W}}\} \) commute, in the algebra of linear
operators on \( \AA \). But from (\ref{003}), (\ref{103}):
\bq
\{{H}\}.\{{\RR {W}}\} - \{{\RR {W}}\}.\{{H}\} = \{\{{H}\}\RR {W}\} = 0
\label{036}
\eq
The formula (\ref{025}) is proven. And also the theorems 1 and 2 which
are only some rewrittings of it. \QED

\section{- The 2 Main Applications}
\label{S03}
\setcounter{equation}{0}
\setcounter{application}{0}
\begin{application} - Classical Mechanics:
\end{application}
Here, an automorphism is called a ``canonical transformation''.

First we have to explicit the Poisson bracket:
\bq
\forall ({A},\Delta) \in \HH \times \bbZ^{L} \label{038}
\eq
\[
(\{{W}\}{V})({A},\Delta) = \ii \sum_{\Delta' \in \bbZ^{L}} \biggl(
{W}'({A},\Delta - \Delta')\cdot\Delta' \biggr). {V}({A},\Delta') -
\biggl( {V}'({A},\Delta - \Delta')\cdot\Delta' \biggr). {W}({A},\Delta')
\]
where \( {W}' \) is the derivative (gradient) of \( {W} \) with
respect to \( {A} \). And we have used a scalar product \( {W}' ({A},
\Delta - \Delta') \cdot \Delta' \). If \( {H} \) is ``non-resonant'',
the action of the operator \( \RR \) on any observable \( {W} \)
consists in keeping (in the Fourier coefficients of \( {W} \)) the
only term with \( \Delta = 0 \), i.e. the average over the angles
variables \( \theta \). So \( \RR {W} \) is a function of \( {A} \)
only (i.e. also ``integrable''), and any 2 functions of \( {A} \)
only, do commute mutually. Then the flow \( e^{{t} \{\RR {W}\}}\ \) is
of the same type as the flow \( e^{{t} \{{H}\}} \), which is:
\bq
e^{{t} \{{H} \}} = e^{{t}. \omega({A}) \partial_{\theta}}
\label{097}
\eq
This is the operator translating the variable \( \theta\ \) by \(\
{t}. \omega({A}) \):
\bq
\bigl( e^{{t} \{{H} \}} {V} \bigr)({A},\theta) = {V}({A},\theta + {t}.
\omega({A})) \label{098}
\eq
or after a Fourier transformation in \( \theta \):
\bq
\bigl( e^{{t} \{{H} \}} {V} \bigr)({A},\Delta) = e^{\ii {t}.
\omega({A}) \cdot \Delta}. {V}({A},\Delta) \label{099}
\eq
And similarly for \( e^{{t} \{\RR {W}\}} \), with \( \omega({A}) \)
replaced by \( (\RR {W})'({A}) = (\RR {W}')({A}) \).

\begin{application} - Quantum Mechanics:
\end{application}
When \( \AA \) has an involution (``*'', named ``adjonction'': cf.
Remark 1), and when \( {H} = {H}^{*} \) the associated automorphism is
called a ``unitary transformation''.

The bracket is defined as:
\bq
\forall {A} \in \HH\ \ \ \ \forall \Delta \in \GG({A}) \label{044}
\eq
\[
(\{{W}\}{V})({A}, \Delta) = {\ii \over \hbar} \sum_{\Delta' \in
\GG({A})} {W}({A} + \Delta', \Delta - \Delta'). {V}({A}, \Delta') -
{V}({A} + \Delta', \Delta - \Delta'). {W}({A}, \Delta')
\]
We can put this bracket in a form similar to the Poisson bracket
(\ref{038}), by adding and substracting 2 terms:
\bq
\forall {A} \in \HH\ \ \forall \Delta \in \GG({A}) \label{046}\\
\eq
\[
(\{{W}\}{V})({A}, \Delta) = \sum_{\Delta' \in \GG({A})} \ii
\Biggl[{{W}({A} + \Delta', \Delta - \Delta') - {W}({A}, \Delta -
\Delta') \over \hbar} \Biggr]. {V}({A}, \Delta') -
\]
\[
\ \ \ \ \ \ \ \ \ \ \ii \Biggl[{{V}({A} + \Delta', \Delta - \Delta') -
{V}({A}, \Delta - \Delta') \over \hbar} \Biggr]. {W}({A}, \Delta') +
\{{W}({A},\Delta')\}{V}({A},\Delta - \Delta')
\]
where the last term is a short notation for:
\bq
\ii \Biggl[ {{W}({A},\Delta').{V}({A},\Delta - \Delta') -
{V}({A},\Delta - \Delta').{W}({A},\Delta') \over \hbar} \Biggr]
\label{047}
\eq
The formula (\ref{046}) is reminiscent of the famous ``correspondance
principle'' between the classical mechanics and the quantum mechanics:
if the set \( \HH \) becomes more and more dense in \( \bbR \) (for
instance if it is \( \hbar \bbZ \)) with \( \hbar \to 0 \), in such a
way that \( {A} \) remains constant, then the last term (\ref{047})
tends to 0 and the first term in bracket tends to the derivative of \(
{W} \) with respect to \( {A} \), multiplied by \( \Delta' \). So that
the quantum bracket becomes the classical one. Cf. \cite{07} for a
precise formulation of this fact, in some particular cases.

From now on, we will choose \( \hbar = 1 \), since we will not use
this semi-classical limit.

Here again, if \( {H} \) is ``non-resonant'', \( \RR {W} \) is a
diagonal matrix, and so its flow is of the same type as the flow of
\( {H} \), since:
\bq
\bigl( e^{t \{{H} \}} {V} \bigr)({A},\Delta) = e^{\ii t. [{h}({A} +
\Delta) - {h}({A})]}. {V}({A},\Delta) \label{101}
\eq
and similarly for the other flow.

\section{- Quantitative Estimates}
\label{S04}
\setcounter{equation}{0}
We start by chosing an arbitrary norm on \( \AA \). And we replace \(
\AA \) by its closure with respect to this norm. We deduce a canonical
norm for the operator \( \Gamma \) which acts bilinearly on \( \AA \):
\bq
|||\Gamma|||:= \sup_{{V},{W} \in \AA\ \st\ ||{V}|| = ||{W}|| = 1} ||
\{\Gamma {W}\} {V}|| \label{052}
\eq

\vspace{0.2cm}
We make an important assumption:
\begin{hypothesis}
\bq
|||\Gamma||| < \infty \label{053}
\eq
\end{hypothesis}
This hypothesis is necessary to be able to apply the so-called ``local
bijection theorem'' to invert the function \( {F} \). When (\ref{053})
is not true, it can be replaced by a weaker one, but then we need to
use the more complicated theorem of Nash-Moser, cf \cite{17},
\cite{18}, \cite{16}, \cite{14}, or the Newton iterative method, as in
the KAM theory. They are based on a Frechet structure on \( \AA\ \)
i.e. an infinite sequence of norms \( ||\ldots||_{{s} - 1} \geq
||\ldots||_{s} \) instead of only 1 norm. The hypothesis (\ref{053})
is still required but with a weaker norm:
\bq
|||\Gamma|||_{\alpha} := \sup_{{s} \in \bbN}\ \ \sup_{{V},{W} \in \AA\
\st\ ||{V}||_{{s} - 1} = ||{W}||_{s} = 1} || \{\Gamma {V}\} {W}||_{s}\
.\ \alpha({s}) \label{102}
\eq
where \( \alpha: \bbN \to \bbR^{*}_{+}\ \ \And\ \ \lim_{{s} \to
\infty}\ \alpha({s}) = 0 \). Hence \( \Gamma \) may be bounded if we
admit some loss of ``regularity''.

Let us also define a norm:
\bq
||||\RR|||| := \sup_{{W} \in \AA\ \st\ ||{W}|| = 1} ||\RR {W}||
\label{106}
\eq
Another assumption, is on the operator \( \RR \):
\begin{hypothesis}
\bq
||||\RR|||| \leq 1 \label{056}
\eq
\end{hypothesis}
It is fulfiled for many norms, for instance those given in
(\ref{051}): cf (\ref{020}).

\vspace{0.2cm}
Then \( {F}\ \) is invertible when \(\ ||{V}||\ \) is small enough:
\vspace{0.2cm}

\hrule
\begin{theorem} -
Let \( {V} \) be an element of \(\ \AA\ \) and \(\ \Gamma\ \) be
defined by (\ref{121}), or explicitly by (\ref{021}) or (\ref{019}).
Under the hypothesis 1, 2 and 3:
\bQ
&& \If\ \ \ \ ||{V}||\ \leq\ {1 \over 5 |||\Gamma|||} \label{054}\\
&& \Then\ \ \ \ {24 \over 35}\ \leq\ {||{F}^{-1}({V})|| \over ||{V}||}
\ \leq\ {24 \over 13} \label{055}
\eQ
\hrule
\end{theorem}

\textbf{Proof}: Let us start by expanding \( {f}({W}) \) as given by
(\ref{009}) in series:
\bq
{f}({W}) = \sum_{n = 1}^{\infty} \{-\Gamma {W}\}^{n} \cdot {n \RR + 1
\over n + 1!} \cdot {W} \label{057}
\eq
and use the definition (\ref{052}):
\bq
||\{\Gamma {W}\} {V}|| \leq |||\Gamma|||.||{W}||.||{V}||
\label{058}
\eq
and the hypothesis (\ref{056}), so that:
\bq
||{f}({W})|| \leq \sum_{n = 1}^{\infty} (|||\Gamma|||.||{W}||)^{n}
\cdot {n ||||\RR|||| + 1 \over n + 1!} \cdot ||{W}|| \leq (e^{
|||\Gamma|||. ||{W}||} - 1).||{W}|| \label{059}
\eq
This proves that \( {f}({W}) = \OO({W}^{2}) \).
To compute the derivative of \( {f}({W}) \) with respect to \( {W} \)
we need the following formula, valid for any derivation \( \partial\
\in\ \Der(\LL)\ \) of some algebra \(\ \LL \):
\bq
\partial e^{V} = \int_{0}^{1} d{t}. e^{{t}.{V}}. \partial {V}. e^{(1 -
{t}).{V}} \label{086}
\eq
which may be rewritten as:
\bq
\partial e^{V} = \Biggl( {e^{\{{V}\}} - 1 \over \{{V}\}} \partial {V}
\Biggr). e^{V} \label{085}
\eq
where the Lie-bracket \(\ \{\ldots\}\ \) is given by the commutator.
Indeed:
\bq
e^{V}. {W}. e^{-{V}} = e^{\{{V}\}} {W}
\label{087}
\eq
The proof of (\ref{085}) is obtained by power expanding the
exponentials:
\bq
\forall {n} \in \bbN\ \ \ \partial({V}^{n}) = \sum_{{k} = 1}^{n} {{n}
\choose {k}}. \{{V}\}^{{k} - 1}. (\partial {V}). {V}^{{n} - {k}}
\label{150}
\eq
which is proven by a simple recurrence. Similarly for (\ref{086}):
\bq
\partial e^{V} = \sum_{{N} = 0}^{\infty} {\partial ({V}^{N}) \over
{N}!} = \sum_{{n}, {k} = 0}^{\infty} {{V}^{n} . \partial {V}. {V}^{k}
\over ({n} + {k} + 1)!} \label{174}
\eq
whereas:
\bq
\int_{0}^{1} d{t}. e^{{t}.{V}}. \partial {V}. e^{(1 - {t}).{V}} =
\sum_{{n}, {k} = 0}^{\infty} \int_{0}^{1} d{t}. {{t}^{n}. (1 -
{t})^{k} \over {n}!. {k}!} {V}^{n}. \partial {V}. {V}^{k} \label{175}
\eq
These 2 expressions coincide after using:
\bq
\int_{0}^{1} d{t}. {{t}^{n}. (1 - {t})^{k} \over {n}!. {k}!} = {1
\over ({n} + {k} + 1)!} \label{176}
\eq
Let us note that (\ref{086}) is a generalization of the following
formula (valid when \( t \) varies in a finite set) to the case where
\( t \) is a continuous variable:
\bq
\partial \Biggl( \prod_{t} {V}_{t} \Biggr) = \sum_{t} \Biggl(
\prod_{\tau < t} {V}_{\tau} \Biggr) (\partial V_{t}) \Biggl(
\prod_{\tau > t} {V}_{\tau} \Biggr) \label{016}
\eq
When \( \partial {V} \) commute with \( {V} \) we retrieve the usual
formula:
\bq
\partial e^{V} = (\partial {V}). e^{V} = e^{V}. (\partial {V})
\label{091}
\eq
Let us apply (\ref{086}) to the Lie algebra \( \LL = \LL(\AA) \) of
the endomorphisms of \( \AA \) (the space of observables), for which
the bracket is indeed the commutator. And we take for \( \partial \)
the derivation with respect to \( {W} \):
\bq
\partial e^{\{ \Gamma {W} \}} = \int_{0}^{1} dt. e^{t.\{ \Gamma {W}
\}}. \{ \Gamma \ldots \}. e^{(1 - t).\{ \Gamma {W} \}} \label{088}
\eq
which can be rewritten as:
\bq
\partial e^{\{ \Gamma {W} \}} = \Biggl( {e^{\{\{ \Gamma {W} \}\}} - 1
\over \{\{ \Gamma {W} \}\}}. \{ \Gamma \ldots \} \Biggr). e^{\{ \Gamma
{W} \}} \label{094}
\eq
where the double bracket is the bracket in \( \LL(\AA) \) i.e. the
commutator:
\bq
\forall X \in \LL(\AA)\ \ \ \ \{\{ \Gamma {W} \}\} X := \{ \Gamma
{W} \}. X - X. \{ \Gamma {W} \} \label{095}
\eq

Therefore:
\bQ
&& {F}'({W}) {V} = \Biggl( \psi \Bigl( \{\{ \Gamma {W} \}\} \Bigr)
\{\Gamma {V}\} \Biggr) e^{-\{ \Gamma {W} \}} \RR {W} + e^{-\{ \Gamma
{W} \}} \RR {V} + \non\\
&&\ \ \ \ \ \ \ \ \ \ \ \Biggl({1 - e^{-\{ \Gamma {W} \}} \over \{
\Gamma {W}\}}\Biggr) \NN {V} + \Biggl( \varphi \Bigl( \{\{ \Gamma {W}
\}\} \Bigr) \{\Gamma {V}\} \Biggr) e^{-\{ \Gamma {W} \}} \NN {W}
\label{092}
\eQ
where:
\bq
\psi (X) := {1 - e^{-X} \over X}\ \ \ \ \ \ \ \ \ \ \ \varphi (X) :=
\int_{0}^{1} ds. \psi (s X) \label{093}
\eq
Let us now use the formula (\ref{058}):
\bq
||e^{\{ \Gamma {W} \}}.{V}|| \leq e^{|||\Gamma|||. ||{W}||}. ||{V}||
\label{090}
\eq
so that:
\bq
||\partial e^{\{ \Gamma {W} \}}.{V}|| \leq \int_{0}^{1} dt. e^{t.
|||\Gamma|||. ||{W}||}. |||\Gamma|||. e^{(1 - t). |||\Gamma|||.
||{W}||}. ||{V}||\ =\ e^{|||\Gamma|||. ||{W}||}. |||\Gamma|||.
||{V}|| \label{089}
\eq
And we don't need the formula (\ref{092}). Hence \( {f}'({W}) \) has a
norm bounded by:
\bQ
&& ||{f}'({W})||\ \leq\ \sum_{n = 1}^{\infty} n. (|||\Gamma|||.||{W}||
)^{n-1}. |||\Gamma||| \cdot {n ||||\RR|||| + 1 \over n + 1!} \cdot
||{W}|| + \non\\
&&\ \ \ \ \ \ \ \ \ \ \ \ \ \ \ \ \ \ (|||\Gamma|||.||{W}||)^{n}\cdot
{n ||||\RR|||| + 1 \over n + 1!} \label{060}
\eQ
so that, using (\ref{056}):
\bq
||{f}'({W})|| \leq e^{|||\Gamma|||.||{W}||}.(|||\Gamma|||.||{W}|| + 1)
- 1 \label{061}
\eq
Let us call \( \gamma \) the solution of the transcendental equation:
\bq
e^{\gamma}(\gamma + 1) - 1 = 1\ \ \ \ie\ \ \ \gamma =
0.3748225258118948\ldots > 13/35 \label{062}
\eq
Hence:
\bq
\If\ \ \ |||\Gamma|||.||{W}|| < \gamma\ \ \ \Then\ \ \ ||{f}'({W})|| <
1 \label{063}
\eq
and so:
\bq
||{F}'({W})|| = ||1 + {f}'({W})|| \geq 1 - ||{f}'({W})|| > 0
\label{064}
\eq
Then \( {F} \) is invertible and:
\bQ
&& ||{F}({W})|| = ||{W} + {f}({W})|| \leq ||{W}||. \Biggl(1 +
{||{f}({W}) || \over ||{W}||} \Biggr) \non\\
&&\ \ \ \ \ \ \ \ \ \ \ \ \ \ \ \ \ \leq ||{W}||. (1 + e^{\gamma} - 1)
= e^{\gamma}. ||{W}|| \label{065}
\eQ
Likewise:
\bq
||{F}({W})|| \geq ||{W}||. \Biggl(1 - {||{f}({W}) || \over ||{W}||}
\Biggr) \geq ||{W}||. (2 - e^{\gamma}) \label{066}
\eq
So if we replace \( {W} \) by \( {F}^{-1}({V}) \) we get:
\bq
e^{-\gamma} \leq {||{F}^{-1}({V})|| \over ||{V}||} \leq {1 \over 2 -
e^{\gamma}} \label{067}
\eq
under the condition (\ref{063}), i.e. if:
\bq
||{V}|| < {(2 - e^{\gamma}).\gamma \over |||\Gamma|||}
\label{068}
\eq
Indeed we will have in that case:
\bq
||{W}|| = ||{F}^{-1}({V})|| \leq {||{V}|| \over 2 - e^{\gamma}} <
{\gamma \over |||\Gamma|||} \label{069}
\eq
To prove the theorem 3, there remains to use the value of \(\gamma \):
\(\ \ e^{\gamma} = {2 \over \gamma + 1} < {35 \over 24} \).
\QED

Let us note that under the condition (\ref{054}), the new term (the
``control term'', \( {f}({V}) \)) will be smaller than \( {V} \) (cf.
(\ref{059})):
\bq
\If\ \ \ \ ||{V}|| \leq {1 \over 5 |||\Gamma|||}\ \ \ \ \ \Then\ \ \ \
{||{f}({V})|| \over ||{V}||} \leq\ e^{1/5} - 1\ \ \ \ \Biggl(<\ {2
\over 9}\Biggr) \label{107}
\eq
\section{- A Formal Series for the Inverse of $F$}
\label{S09}
\setcounter{equation}{0}
We want an expansion of \( {W} = {F}^{-1}({V}) \) in powers of \( {V}
\), using the expansion of \( {F}({W}) = {W} + {f}({W}) \) given in
(\ref{057}). For that purpose we can use an extension of the Lagrange
inversion formula which was established to invert a function from \(
\bbC \to \bbC \) in power series. An extension to the case where the
argument is not a number but a function (like here), was given in
\cite{08}, and latter in \cite{22}. We first rewrite the definition of
\( {W} \) as a fixed-point problem:
\bq
{F}({W}) = {V}\ \ \Longleftrightarrow\ \ {W} = {G}({W}) := {V} -
{f}({W}) \label{115}
\eq
A Taylor expansion of \( {G} \) is given by:
\bq
{G}({W}) = \sum_{n \geq 0} \hat{G}({n}) {W}^{n}
\label{116}
\eq
with (cf. \ref{057}):
\bq
\hat{G}(0) := {G}(0) = {V}\ \ \ \And\ \ \ \hat{G}(1) := 0
\label{042}
\eq
\bq
\forall {n} \geq 2\ \ \ \hat{G}({n}) := {{G}^{({n})}(0) \over {n}!} =
\SS (-1)^{n} \{\Gamma \ldots \}^{{n}-1} \cdot {({n} - 1) \RR + 1 \over
n!} \ldots \label{117}
\eq
is an \( {n} \)-linear completely symmetric application from \(
\AA^{n} \) into \( \AA \). Its \( {n} \) arguments are symbolized by
the \( {n} \) ``slots''. In (\ref{116}), this \( {n} \)-linear mapping
is applied to identical arguments: \( {n} \) times \( {W} \). When
applied to \( {n} \) general arguments \( {W}_{1},\ldots,{W}_{n} \),
it would give:
\bq
\hat{G}({n}) ({W}_{1},\ldots,{W}_{n}) = \SS (-1)^{n} \{\Gamma {W}_{1}
\}\ldots \{\Gamma {W}_{{n} - 1} \} \cdot {({n} - 1) \RR + 1 \over n!}
{W}_{n} \label{123}
\eq
The operator \( \SS \) is the symmetrization of the arguments which
yields, when applied to an \( {n} \)-linear application \( {T} \):
\bq
(\SS {T}) ({W}_{1},\ldots,{W}_{n}) := {1 \over n!}\ \sum_{\sigma \in
\mathrm{\,Permutations}} {T}
({W}_{\sigma(1)},\ldots,{W}_{\sigma({n})}) \label{131}
\eq
We will omit the parenthesis and the commas in using such tensors. So
\( {W}^{M} = ({W},{W},\ldots,{W}) \) (M times). For instance:
\bq
{G}'(0)({W}_{1}) := \lim_{\lambda \to 0} {{G}(\lambda {W}_{1}) -
{G}(0) \over \lambda} \label{136}
\eq
which is 0, for the function defined in (\ref{115}), since it is
quadratic in its argument. And \( {G}''(0) \) is a tensor of order 2.
\bq
{G}''(0)({W}_{1}, {W}_{2}) := \lim_{\lambda \to 0} \lim_{\mu \to 0}
{{G}(\lambda {W}_{1} + \mu {W}_{2}) - {G}(\lambda {W}_{1}) - {G}(\mu
{W}_{2}) + {G}(0) \over \lambda. \mu} \label{135}
\eq
which can be easily computed to be:
\bq
\hat{G}(2) {W}_{1}. {W}_{2} = \{\Gamma {W}_{1} \} {\RR + 1 \over 4}
{W}_{2} + \{\Gamma {W}_{2} \} {\RR + 1 \over 4} {W}_{1} \label{132}
\eq

\newpage
\hrule
\begin{theorem} -
The above mentionned extension of the Lagrange inversion formula says
that the solution of any fixed-point equation \( {W} = {G}({W}) \),
with a function \( {G} \) given by a series (\ref{116}), is formally:
\bq
{W} = \sum_{{N} \geq 1} \sum_{\nu \in \TT({N})} \hat{G}(\nu_{{N}-1})
\hat{G}(\nu_{{N}-2})\ldots \hat{G}(\nu_{1}) \hat{G}(\nu_{0})
\label{118}
\eq
where:
\bQ
&& \TT({N}) := \{ \nu = (\nu_{0}, \nu_{1}, \ldots, \nu_{{N} - 1}) \in
\{ 0, 1, \ldots, {{N} - 1} \}^{N}\ \ \st\non\\
&&\ \ \ \ \ \ \ \ \ \ \ \ \ \forall {k} \in \{ 0, 1, \ldots, {{N} - 1}
\}:\ |\nu|_{k} \leq {k}\ \ \And\ \ |\nu|_{{N} - 1} = {{N} - 1} \}
\label{119}
\eQ
with:
\bq
|\nu|_{k} := \nu_{0} + \nu_{1} + \ldots + \nu_{k}
\label{120}
\eq
\hrule
\end{theorem}

The expansion (\ref{118}) is only useful if \( \hat{G}({n}) \) are
``small'', since we expand in powers of them. In our case (\ref{042}),
(\ref{117}), only \( \hat{G}(0) \) is small, and \( \hat{G}(1) = 0 \).
But \( \hat{G}({n}) \) is of order 1, when \( {n} \geq 2 \). So we
have to rearrange the series (\ref{118}):
\bq
{W} = {V} + \sum_{M \geq 2} {W}_{M} \label{133}
\eq
where:
\bq
{W}_{M} := \sum_{{N} = {M} + 1}^{2 {M} - 1}\ \ \sum_{\nu \in \TT({N})\
\st\ \nu^{-1}(0) = {M}}\ \ \hat{G}(\nu_{{N} - 1}) \hat{G} (\nu_{{N} -
2}) \ldots \hat{G} (\nu_{1}) \hat{G}(\nu_{0})
\label{166}
\eq
with the restriction ``\( \nu^{-1}(0) = {M} \)'' in the sum over \(
\nu \), meaning that \( \nu \) must take exactly \( {M} \) times the
value 0. Indeed this is exactly the case ``\( \nu_{k} = 0 \)'' which
produces a factor \( {V} \). We can also notice that in the definition
(\ref{119}) of \( \TT({N}) \) we can restrict \( \nu \in \{ 0, 2, 3,
\ldots, {N} - 1 \}\ \) since in our case \(\ \hat{G}(1) = 0 \). Before
we explicit the first terms of the solution (\ref{133}), let us
understand heuristically the series (\ref{118}) or (\ref{133}). A
first approximation of the solution of \( {W} = {G}({W})\ \) is \(\
{W} \approx {G}(0) := {V} \). Then a second approximation is \(\ {W}
\approx {G}({G}(0)) \approx {G}(0) + {G}'(0).{G}(0) \). And a third
one, at the order \( {V}^{3}\ \), is \(\ {W} \approx {G}({G}({G}(0)))
\approx {G}[{G}(0) + {G}'(0).{G}(0)]\ \) i.e.:
\bq
{W} \approx {G}(0) + {G}'(0). {G}(0) + {G}'(0). {G}'(0). {G}(0) + {1
\over 2}. {G}''(0). {G}(0). {G}(0) \label{134}
\eq
and so on. We are building the series (\ref{118}). Now we have to
check that the general term of (\ref{118}) is indeed an element of \(
\AA \) and not an arbitrary \( {m} \)-linear mapping for some \( {m}
\). Each term is a composition of high-order tensors, and this product
is actually a ``vector'' i.e. an element of \( \AA \). This is due to
the definition of \( \TT({N}) \): we note that \( \hat{G}({n}) \) is
of type \(\ {V}.\bar{V}^{n}\ \) i.e. it is a (sum of) tensorial
product(s) of a vector \( {V} \) and of \( {n} \) covectors \( \bar{V}
\). More generally we say that a tensor is of type \( {V}^{m}.
\bar{V}^{n}\ \) when it is \( {n} \) times covariant and \( {m} \)
times contravariant. This is a convenient way to consider such
tensors, in order to keep track of all subsequent contractions. Of
course the ordering is crucial: \( {V}. \bar{V} \) is a matrix (a
tensor of order 2) whereas \( \bar{V}. {V} \) is a scalar (a scalar
product). Each time a covector follows (on the left) a vector, we make
the contraction, i.e. we reduce the tensorial product to a scalar.
Actually we just want to give a meaning to the composition of
multilinear applications. To this purpose we can define:
\bq
\forall {B} \in \LL(\AA^{n},\AA)\ \ \ \forall {C} \in \LL(\AA^{m},\AA)
\ \ \ {B}. {C} \in \LL(\AA^{{m} + {n} - 1},\AA) \label{124}
\eq
by:
\bq
({B}. {C})({V}_{1}, \ldots, {V}_{m}, {V}_{{m} + 1}, \ldots, {V}_{{m} +
{n} - 1}) := {B} \Bigl( {C} ({V}_{1},\ldots, {V}_{{m}}), {V}_{{m} +
1},\ldots, {V}_{{m} + {n} - 1} \Bigr) \in \AA \label{125}
\eq
Now we note that for any \( \nu \in \TT({N}): \nu_{0} = 0 \) since \(
|\nu|_{0} = 0 \). So the right-most factor in (\ref{118}) is a vector:
\( \hat{G}(0) = {V} \in \AA \).

Then the preceding one is \( \hat{G}(\nu_{1})\ \) where \(\ \nu_{1}
\leq 1\ \) since \(\ |\nu|_{1} = \nu_{0} + \nu_{1}\ \) has to be less
than 1. Hence \( \hat{G}(\nu_{1})\ \) is either a vector (if \(
\nu_{1} = 0 \)) or a matrix (if \( \nu_{1} = 1 \)). In the latter case
the product \(\ \hat{G}(1) \hat{G}(0)\ \) is a vector, and in the
former case we have a ``bi-vector'' which will be made into a vector
by the higher-order terms: indeed if \(\ \nu_{1} = 0\ \) then there
exists a \(\ {k} \geq 2\ \) such that \( \nu_{k} \geq 2\ \) since \(\
|\nu|_{{N} - 1}\ \) has to be \(\ {N} - 1 \). More precisely we note
that the product of a tensor \(\ {V}^{a}. \bar{V}^{b}\ \) by a tensor
\(\ {V}^{c}. \bar{V}^{d}\ \) is a tensor of type:
\bq
{V}^{a}. \bar{V}^{b}. {V}^{c}. \bar{V}^{d} = {V}^{{a} + {c} -
\min({b}, {c})}. \bar{V}^{{b} + {d} - \min({b}, {c})} \label{137}
\eq
So the product \(\ \hat{G}(\nu_{k}) \ldots \hat{G}(\nu_{0})\ \) is a
tensor of type:
\bq
\hat{G}(\nu_{k}) \ldots \hat{G}(\nu_{0}) = {V}^{\max({k} -
|\nu|_{k},1)}. \bar{V}^{0} \label{138}
\eq
i.e. without any covector, since we assume \( |\nu|_{k} \leq {k}\ \
\forall {k} \). Finally for \( {k} = {N} - 1,\ |\nu|_{{N} - 1} \) has
to be \( {N} - 1 \), so (\ref{118}) is indeed a sum of vectors.

The idea of the proof of (\ref{118}) is to plug the r.h.s. of
(\ref{118}) into the expression (\ref{116}) of \( {G}({W}) \) and to
recognise that we get the same expansion than (\ref{118}), i.e. this
is also \( {W} \).

\vspace{0.2cm}
A simple illustration (which is not of our type) of the theorem 4 is
when:
\bq
{G}({W}) := {V}. e^{\bar{Y} \cdot {W}}
\label{017}
\eq
for some \( {V} \in \AA\ \And\ \bar{Y} \in \AA^{*} := \LL(\AA, \bbR)
\). In that simple case, the theorem 4 gives the solution of the
``transcendental'' equation in \( {W} \):
\bq
{W} = {V}. e^{\bar{Y} \cdot {W}}
\label{126}
\eq
as:
\bq
{W} = {V}. \lambda(\bar{Y} \cdot {V})\ \ \ \where\ \ \ \lambda({x}) :=
\sum_{{n} \geq 0} {({n} + 1)^{{n} - 1} \over n!} {x}^{n} \label{127}
\eq
So that \( \lambda({x}) \) converges if \( |{x}| \leq e^{-1} \) (but
may be extended if \( {x} < - e^{-1} \)).

\vspace{0.2cm}
We can rearrange the series (\ref{133}) as follows:
\bq
{W}_{M} := \sum_{{N} = 1}^{{M} - 1}\ \ \sum_{\lambda \in \BB({N},{M})}
\ \sum_{\mu \in \CC({N}, \lambda)} \hat{G}(\lambda_{1}^{\mu}) \ldots
\hat{G}(\lambda_{{N} + {M}}^{\mu}) \label{167}
\eq
where \( \hat{G} \) is defined in (\ref{042}) and (\ref{117}), and:
\bq
\BB({N},{M}) := \{ \lambda \in \bbN_{*}^{N}\ \ \st\ \ \ |\lambda|_{N}
= {M} - 1 \} \label{168}
\eq
and:
\bq
\CC({N}, \lambda) := \{ \mu \in \bbN_{*}^{N}\ \ \st\ \ \ \forall\ 1
\leq {n} \leq {N}\ \ \ \ |\mu|_{n} \leq {n} + |\lambda|_{{n} - 1} \}
\label{169}
\eq
with:
\bq
|\mu|_{n} := \mu_{1} + \ldots + \mu_{n}
\label{170}
\eq
ans similarly for \( |\lambda|_{n} \) with \( |\lambda|_{0} := 0 \).
Finally in (\ref{167}) the integer \( \lambda_{l}^{\mu} \) is defined
for any \( 1 \leq {l} \leq {N} + {M} \) by:
\bQ
\lambda_{l}^{\mu} &=& \lambda_{n} + 1\ \ \ \mathrm{if}\ \ \ {l} =
|\mu|_{n} \non\\
&=& 0 \ \ \ \ \mathrm{otherwise} \label{171}
\eQ
This means that \( \lambda_{l}^{\mu} = 0 \) when \( {l} \) is not an
element of the set of the values \( |\mu|_{n} \), when \( {n} \)
varies in \( \{ 1, \ldots ,{N} \} \). This case occurs \( {M} \) times
when \( {l} \) varies in \( \{ 1, \ldots ,{N} + {M} \} \). Otherwise
when \(\ {l} = |\mu|_{n}\ \) for some \( {n} \), then \(
\lambda_{l}^{\mu} = \lambda_{n} + 1 \), for this \( {n} \).

\vspace{0.2cm}
Finally we can also rearrange the Lagrange series in our specific case
(\ref{042}) \& (\ref{117}) into:
\bq
{W}_{M} := \sum_{{N} = 1}^{{M} - 1}\ \ \sum_{\lambda \in \BB({N},{M})}
\ \sum_{\nu \in \DD({N}, \lambda)} (-1)^{{N} + {M} + 1} \{ \Gamma
\ldots \}^{\lambda_{N}} {\lambda_{N} \RR + 1 \over \lambda_{N} + 1!}
{V}^{\nu_{N}} \cdot \ldots \cdot \{ \Gamma \ldots \}^{\lambda_{1}}
{\lambda_{1} \RR + 1 \over \lambda_{1} + 1!} {V}^{\nu_{1}} \label{177}
\eq
where:
\bq
\DD({N}, \lambda) := \{ \nu \in \bbN^{N}\ \ \st\ \ \ |\nu|_{N} = {M}\
\ \ \&\ \ \ \forall\ 1 \leq {n} \leq {N}\ \ \ \ |\nu|_{n} >
|\lambda|_{{n}} \} \label{007}
\eq
Let us note that:
\bq
||\hat{G}({n})|| \leq {|||\Gamma|||^{{n} - 1} \over ({n} - 1)!}
\label{172}
\eq
and so:
\bq
||{W}_{M}|| \leq {c}_{M}. |||\Gamma|||^{{M} - 1}. ||{V}||^{M}
\label{173}
\eq
for some positive constant \( {c}_{M} \). Indeed: \( \lambda_{1} +
\ldots + \lambda_{N} = {M} - 1 \).

\vspace{0.2cm}
Let us explicit the first orders of the expansion of the solution \(
{W} \) of our problem (\ref{115})-(\ref{117}), as given by
(\ref{133}) or (\ref{167}):
\bq
{F}^{-1}({V}) = {W} = \sum_{{M} \geq 1} {W}_{M}\ \ \ \ \ \where\ \ \
{W}_{1} = {V}\ \ \ \ \ \ {W}_{2} = \{\Gamma {V} \} {\RR + 1 \over 2}
{V} \label{139}
\eq
\[
{W}_{3} = \Biggl( \{\Gamma {V} \} {\RR + 1 \over 2} \Biggr)^{2} {V} -
\{\Gamma {V} \}^{2} {2 \RR + 1 \over 6} {V} + \Biggl\{ \Gamma \biggl(
\{\Gamma {V} \} {\RR + 1 \over 2} {V} \biggr) \Biggr\} {\RR + 1 \over
2} {V}
\]
\[
{W}_{4} = \Biggl( \{\Gamma {V} \} {\RR + 1 \over 2} \Biggr)^{3} {V} +
\{\Gamma {V} \}^{3} {3\RR + 1 \over 24} {V} - \{\Gamma {V} \}^{2}
{2\RR + 1 \over 6} \{\Gamma {V} \} {\RR + 1 \over 2} {V} -
\]
\[
\Biggl\{ \Gamma \biggl( \{\Gamma {V} \} {\RR + 1 \over 2} {V} \biggr)
\Biggr\} {\RR + 1 \over 2} \{\Gamma {V} \} {\RR + 1 \over 2} {V} -
\Biggl\{ \Gamma \biggl( \{\Gamma {V} \}^{2} {2\RR + 1 \over 6} {V}
\biggr) \Biggr\} {\RR + 1 \over 2} {V} -
\]
\[
\{\Gamma {V} \} {\RR + 1 \over 2} \{\Gamma {V} \}^{2} {2\RR + 1 \over
6} {V} + \{\Gamma {V} \} {\RR + 1 \over 2} \Biggl\{ \Gamma \biggl(
\{\Gamma {V} \} {\RR + 1 \over 2} {V} \biggr) \Biggr\} {\RR + 1 \over
2} {V} +
\]
\[
\Biggl\{ \Gamma \Biggl( \Biggl\{ \Gamma \biggl( \{\Gamma {V} \} {\RR +
1 \over 2} {V} \Biggr) \Biggr\} {\RR + 1 \over 2} {V} \biggr) \Biggr\}
{\RR + 1 \over 2} {V} + \Biggl\{ \Gamma \Biggl( \Biggl( \{\Gamma {V}
\} {\RR + 1 \over 2} \Biggr)^{2} {V} \Biggr) \Biggr\} {\RR + 1 \over
2} {V} -
\]
\[
\Biggl\{ \{\Gamma {V} \} \Gamma \biggl( \{\Gamma {V} \} {\RR + 1 \over
2} {V} \biggr) \Biggr\} {2\RR + 1 \over 6} {V}
\]
and so on for \( {W}_{5} \)\ldots We have proven in Theorem 3 that
this series converges at least when:
\bq
||{V}||\ \leq\ {1 \over 5 |||\Gamma|||} \label{128}
\eq

\vspace{0.2cm}
The paper \cite{23} proves that a series similar to this one (the
Lindstedt series, which is local in the variable \( {A} \)) converges.
The key ingredient is the compensation between the terms of different
signs.

\vspace{0.2cm}
Remark: as we have noticed at the end of section \ref{S01}, the
solution \( {W} \) is independant of the constant \(\ \hbar\ \) since
the operator \(\ \Gamma\ \) always appears with a bracket around it.
And the pair \( \{ \Gamma \ldots \} \) is independant of \(\ \hbar\
\).

\section{- Example 1: Classical Control Theory}
\label{S07}
\setcounter{equation}{0}
We will study a simple model, introduced in \cite{19} and described in
\cite{15}, \cite{25}, of a charged particle in a plasma, in a Tokamak,
which is a reactor for the controlled thermonuclear fusion. We
consider a section of the Tokamak as the phase space of a dynamical
system with 1 degree of freedom. So that the particle has 1 degree of
freedom \( ({p},{q}) \), but it is embedded in a complicated electric
field, depending on time. The fast motion given by the strong magnetic
field has been averaged out. The hamiltonian is:
\bQ
&& {H}({p},{q},{E},{\tau}) + {V}({p},{q},{E},{\tau})\ \ \ \where\ \ \
{H}({p},{q},{E},{\tau}) = {E} \label{074}\\
&& \And\ \ \ {V}({p},{q},{E},{\tau}) = \sum_{{n}, {m}, {k} \neq 0}
\varepsilon^{(1)}_{{n}, {m}, {k}}. \sin({n} {q} + {m} {p} + {k}
{\tau}) \label{075}
\eQ
The extended canonical coordinates are \( ({E},\tau) \) so that the
motion of the new dynamical variable \( \tau \) is trivial:
\bq
e^{{t}.\{{H} + {V}\}} \tau = \tau + {t}
\label{076}
\eq
Here we have applied the flow to the observable \( \tau: ({p}, {q},
{E}, {\tau}) \mapsto \tau \). And of course \( {V} \) is independant
of the variable \( {E} \), which is the variable canonically
conjugate to \( \tau \). Let us note that in (\ref{075}), the
variable \( {k} \) must be integer (or at least away from 0) but the
variables \( {m}, {n} \) may be integer or real numbers: in that case
the sum over them should be replaced by an integral.

The hamiltonian \( {H} \) is resonant since \( \{{H}\} =
\partial_{\tau} \) so that:
\bq
(\RR {V})({p},{q},\tau) = \oint d\tau. {V}({p},{q},\tau)
\label{104}
\eq
is independent of \( \tau \). Indeed a Fourier transformation in \(
\tau \) yields:
\bq
\hat{(\RR {V})}({p}, {q}, {E}, {k}) = \hat{V}({p}, {q}, {E}, 0).\
\chi( {k} = 0) \label{165}
\eq
For the perturbation (\ref{075}) we have: \( \RR {V} = 0 \). The
action of the operator \( \Gamma \) is defined by:
\bq
(\Gamma {V})({p},{q},{\tau}) = -\sum_{{n}, {m}, {k} \neq 0}
{\varepsilon^{(1)}_{{n}, {m}, {k}} \over {k}} \cdot \cos({n} {q} + {m}
{p} + {k} {\tau}) \label{077}
\eq
Then the ``control term'' \( {f}({V}) \) can be explicitly computed,
cf (\ref{057}) with \( \RR {V} = 0 \):
\bq
{f}({V}) = \sum_{s \geq 2} {f}_{s}\ \ \ \ \where\ \ \ \ {f}_{s} :=
{\{-\Gamma {V}\}^{s - 1} \over s!} {V} \label{105}
\eq
Hence:
\bq
{f}({V})({p},{q},{\tau}) = \sum_{{n}, {m}, {k} \neq 0}
\varepsilon_{{n}, {m}, {k}}. \sin({n} {q} + {m} {p} + {k}
{\tau}) \label{078}
\eq
where:
\bq
\varepsilon_{{n}, {m}, {k}}:= \sum_{s \geq 2} (-1)^{s-1} \cdot
{\varepsilon^{(s)}_{{n}, {m}, {k}} \over s!} \label{079}
\eq
with:
\bq
\varepsilon^{(s)}_{{n}, {m}, {k}}:= \sum_{{N}, {M}, {K} \neq 0}
{\varepsilon^{(1)}_{{N}, {M}, {K}} \over {K}} \cdot \biggl(
\varepsilon^{(s - 1)}_{{N + n}, {M + m}, {K + k}} - \varepsilon^{(s -
1)}_{{N - n}, {M - m}, {K - k}} \biggr).({M}. {n} - {N}. {m})
\label{080}
\eq
The proof of (\ref{080}) is based on:

\[
\hspace{-6cm}\Bigl\{ \cos({N} {q} + {M} {p} + {K} {\tau}) \Bigr\}
\sin({n} {q} + {m} {p} + {k} {\tau}) =
\]
\[
\ \ \ \ \ \ \ \ \ \ \ \ \ \ \ \ {1 \over 2} \cdot ({N}. {m} - {M}.
{n}) \cdot \biggl[ \sin[({N} + {n}) {q} + ({M} + {m}) {p} + ({K} +
{k}) \tau]\ +
\]
\bq
\ \ \ \ \ \ \ \ \ \ \ \ \ \ \ \ \ \ \ \ \ \ \ \ \ \ \ \ \sin[({N} -
{n}) {q} + ({M} - {m}) {p} + ({K} - {k}) \tau] \biggr] \label{122}
\eq
so that we can iterate and compute \( \{\Gamma {V}\}^{s}\ {V} \).

\vspace{0.2cm}
A simple case where the control term can be computed explicitly is the
following. Let us choose:
\bq
\varepsilon \in \bbR^{*},\ \ {b} > 1/\sqrt{2},\ \ \sigma = \pm 1,\ \
{m}, {n} \in \bbR\ \ \ \st\ \ {m} \neq {n} \label{110}
\eq
And we take \( {V} \) as a particular case of (\ref{075}), a sum of 2
waves:
\bq
{V}({p},{q},{\tau}) = \varepsilon. \sigma. \sqrt{2 {b}^{2} - 1}.
\sin({n} {q} + {m} {p} + \tau) - \varepsilon. \sin({q} + {p} + \tau)
\label{111}
\eq
We have taken \( {m} \neq {n} \) to avoid that \( {V} \) depends only
on a single variable \( {q} + {p} \). The role of \( {b} \) (and \(
\sigma \)) is to permit 2 different coupling constants. Then the
control term is a sum of only 5 waves:
\bQ
&& {f}({V}) = \hat{\varepsilon}^{2}. \sigma. \sin\Bigl( ({n} - 1) {q}
+ ({m} - 1) {p} \Bigr) + \non\\
&& \ \ \ \ \ \tilde{\varepsilon}^{3}. \Biggl[ \sqrt{2 {b}^{2} - 1}.
\Biggl( \sin({q} + {p} + \tau) + \sin\Bigl( (2{n} - 1) {q} + (2{m} -
1) {p} + \tau \Bigr) \Biggr) - \non\\
&& \ \ \ \ \ \sigma. \Biggl( \sin({n} {q} + {m} {p} + \tau) +
\sin\Bigl( (2 - {n}) {q} + (2 - {m}) {p} + \tau \Bigr) \Biggr) \Biggr]
\label{112}
\eQ
where:
\bq
\hat{\varepsilon} := {(2 {b}^{2} - 1)^{1/4} \over {b}}\cdot \Biggl( {1
- \cos(\varepsilon. {b}. ({m} - {n})) \over |{m} - {n}|} \Biggr)^{1/2}
\approx |\varepsilon|. |{m} - {n}|^{1/2}. \Bigl( {2 {b}^{2} - 1 \over
4} \Bigr)^{1/4} \label{113}
\eq
\bq
\tilde{\varepsilon} := {(4 {b}^{2} - 2)^{1/6} \over {b}}\cdot \Biggl(
\varepsilon. {b} - {\sin(\varepsilon. {b}. ({m} - {n})) \over {m} -
{n}} \Biggr)^{1/3} \approx \varepsilon. |{m} - {n}|^{2/3}. \Bigl( {2
{b}^{2} - 1 \over 18} \Bigr)^{1/6} \label{114}
\eq
We have indicated the first order of the expansion of \(\
\hat{\varepsilon}\ \) or \(\ \tilde{\varepsilon}\ \) when \(\
\varepsilon\ \) is small. So that \( {f}({V}) = \OO({V}^{2}) \).

See also \cite{15}, \cite{25} for some numerical experiments that prove
the effectiveness of this method, when the coefficients \(
\varepsilon_{ {n}, {m}, {k}} \) are taken to reflect some properties
of a realistic field, in a Tokamak:
\bq
\varepsilon_{{n}, {m}, {k}} = {\varepsilon \over ({n}^{2} +
{m}^{2})^{3 \over 2}} \cdot \chi(1 \leq {n}^{2} + {m}^{2} \leq
{N}^{2}).\ \chi({k} = 1) \label{140}
\eq
for some constant \( \varepsilon \) proportional to the inverse of the
(strong) magnetic field and for some ``cut-off'' \( {N} \). In that
case the first term of the control, \( \varepsilon_{{n}, {m},
{k}}^{(2)} \) is vanishing when \( {k} \neq 0 \).

In \cite{25} we also give some quantitative values of the parameters
for the rigorous applicability of this control.

\vspace{0.2cm}
Let us summarize the method of control, in this case, where \( \RR {V}
= 0 \) and \( \{{H}\} = \partial_{\tau} \):
\bq
\forall {t} \in \bbR\ \ \ \ \ e^{{t} \{{H} + {V} + {f}({V})\}} =
e^{-\{\Gamma {V}\}}. e^{{t} \partial_{\tau}}. e^{\{\Gamma
{V}\}} \label{141}
\eq
so that the distance between the dynamical variable \( {p}\ \) (or \(\
{q}\ \)) at the time \( {t} \) and at the initial time is:
\bq
{p}_{t} - {p}_{0} = \biggl( e^{{t} \{{H} + {V} + {f}({V})\}} - 1
\biggr) {p}_{0} \label{142}
\eq
Let us write \( {p} \) instead of \( {p}_{0} \), and use the
``telescopic'' formula:
\bq
\forall a, b, c:\ \ \ a. b. c - 1 = (a - 1) + a.(b - 1) + a. b. (c -
1) \label{143}
\eq
And we replace the flow in (\ref{142}) by its decomposition
(\ref{141}):
\bq
{p}_{t} - {p} = \Biggl[ (e^{-\{\Gamma {V}\}} - 1) + e^{-\{\Gamma
{V}\}}. (e^{{t} \partial_{\tau}} - 1) + e^{-\{\Gamma {V}\}}. e^{{t}
\partial_{\tau}}. (e^{\{\Gamma {V}\}} - 1) \Biggr] {p} \label{144}
\eq
But the middle term vanishes since \( \partial_{\tau} {p} = 0 \), so
that:
\bq
{p}_{t} - {p} = (e^{-\{\Gamma {V}\}} - 1) {p} + e^{-\{\Gamma
{V}\}}. e^{{t} \partial_{\tau}}. (e^{\{\Gamma {V}\}} - 1) {p}
\label{145}
\eq
And we can divide and multiply by \( \{\Gamma {V}\} \), and use the
antisymmetry (\ref{002}):
\bq
{p}_{t} - {p} = \biggl( {1 - e^{-\{\Gamma {V}\}} \over \{\Gamma
{V}\}} \biggr) \{{p}\} \Gamma {V} - e^{-\{\Gamma {V}\}}. e^{{t}
\partial_{\tau}}. \biggl( {e^{\{\Gamma {V}\}} - 1 \over \{\Gamma
{V}\}} \biggr) \{{p}\} \Gamma {V} \label{146}
\eq
Finally, let us note that \( \{{p}\} = -\partial_{q} \), i.e. the
formula (\ref{146}) can be explicitly computed.

\vspace{0.2cm}
When we apply an approximate control term \( \varphi \) instead of the
exact one \( {f}({V}) \), the formula (\ref{141}) becomes:
\bq
e^{{t} \{{H} + {V} + \varphi\}} = e^{-\{\Gamma {V}\}}. e^{-\{\Gamma
{W}\}}. e^{{t} \partial_{\tau}}. e^{{t} \{\RR {W}\}}. e^{\{\Gamma
{W}\}}. e^{\{\Gamma {V}\}} \label{147}
\eq
where:
\bq
{W} := {F}^{-1} \biggl( e^{\{\Gamma {V}\}}. \Bigl( \varphi - {f}({V})
\Bigr) \biggr) \label{148}
\eq
When we take \( \varphi = 0 \), (\ref{148}) is the begining of the KAM
recursive method. See (\cite{30}) for some extensions and numerical
tests of this theory for some dynamical systems. See also \cite{14}
for more details, in the case of quantum mechanics.

\section{- Example 2: Quantum Adiabatic Transformation}
\label{S06}
\setcounter{equation}{0}
An example of typical norm on \( \AA \) is given by an arbitrary
``weight'' function \(\ {g}({A}, \Delta) > 0 \). Let us define a norm
on the Lie-algebra \( \AA \) by:
\bq
||{V}||:= \sup_{{A}} \sum_{\Delta} |{V}({A},\Delta)|/{g}({A},\Delta)
\label{051}
\eq
In the quantum mechanical case, we will take the usual \( L^{2}
\)-operator norm on each ``block'' \( {V}({A},\Delta) \). This choice
is irrelevant in the case where all the projectors \( {P}_{A} \) are
finite-dimensional. We could have chosen \(\ \ \sup_{{A}, \Delta}\ \)
but this would have just been (approximately) a modification of the
``weight'' function \( {g} \).

\begin{lemma}
We have:
\bq
||\RR {V}|| \leq ||{V}|| \ \ \ \ \&\ \ \ \ ||\NN {V}|| \leq ||{V}||
\label{020}
\eq
\end{lemma}
\textbf{Proof}: The operators \( \RR \) and \( \NN \) are implemented
by some characteristic functions: cf (\ref{041}), (\ref{039}),
(\ref{050}), (\ref{049}). So the sup \& sum in (\ref{051}) are
restricted by some conditions: therefore the norm decreases. \QED

Hence the hypothesis 3 is fulfiled. Let us consider as before a
hamiltonian \( {H} \) but where now the perturbation depends on time.
We need to extend the algebra and the full hamiltonian is now: \( {D}
+ {H} \) where \( {D} \) is the ``derivative with respect to time''.
We can apply the proposition 1 (\ref{028}) but we still use the same
operator \( \Gamma\ \) i.e. the pseudo-inverse of the bracket with \(
{H} \), and also the same function \( {F} \) defined in (\ref{008}).
Let us rewrite (\ref{028}):
\bq
\forall {W} \in \AA\ \ \ \ {H} + {F}({W}) = e^{-\{\Gamma {W}\}} ({H} +
\RR {W})
\eq
Now we want some information on \( {D} + {H} + {V} \):
\bq
{D} + {H} + {F}({W}) = {D} + e^{-\{\Gamma {W}\}} ({H} + \RR {W}) =
e^{-\{\Gamma {W}\}} (e^{\{\Gamma {W}\}}{D} + {H} + \RR {W})
\label{070}
\eq
where:
\bq
{W} := {F}^{-1}({V})
\label{084}
\eq
Hence:
\bq
{D} + {H} + {F}({W}) = e^{-\{\Gamma {W}\}} \Biggl( {D} + {e^{\{\Gamma
{W}\}} - 1 \over \{\Gamma {W}\}}.\{\Gamma {W}\}{D} + {H} + \RR {W}
\Biggr) \label{071}
\eq
so, using (\ref{002}), and the notation \( \dot{W}:= \{{D}\}{W} \):
\bq
{D} + {H} + {V} = e^{-\{\Gamma {W}\}} \Bigl( {D} + {H}_{1} + {V}_{1}
\Bigr) \label{072}
\eq
with:
\bq
{H}_{1} := {H} + \RR {W}\ \ \ \ \ {V}_{1} := - {e^{\{\Gamma {W}\}} - 1
\over \{\Gamma {W}\}}.\Gamma \dot{W} \label{083}
\eq
Indeed \( \{{D}\} \Gamma {W} = \Gamma \{{D}\} {W} = \Gamma \dot{W} \).
The formula (\ref{072}) is useful if the derivative \( \dot{V} \) of
the perturbation with respect to time is ``smaller'' than the
perturbation \( {V} \) itself. This is the ``adiabatic hypothesis''.
In that case we can also show that \( \dot{W} \) is small, since \(
{W} \approx {V} \). So the new perturbation is ``smaller'' than the
original perturbation, or more qualitatively it is approximately: \(
\Gamma \dot{W}\ \) i.e. \(\ \{{H}\}^{-1}\{{D}\}{V} \).
This means that it has losed some regularity in time, but gained some
regularity in its ``spatial'' dependance, since generally \(
\{{H}\}^{-1} \) is a regularizing operator. The difference between \(
{H} \) and the new \( {H}_{1} \) is of the same size as the original
perturbation, but ``diagonalized'' with respect to \( {H} \) since it
commutes with it. Cf also \cite{13} for an iterative proof of this
result.

Let us finally note that it is possible to iterate this procedure:
\bq
{D} + {H} + {V} = e^{-\{\Gamma {W}\}} e^{-\{\Gamma_{1} {W}_{1}\}}
\Bigl( {D} + {H}_{2} + {V}_{2} \Bigr)\ \ \ \ \ \where\ \ \ \ \ {W}_{1}
:= {F}_{1}^{-1}({V}_{1}) \label{108}
\eq
and:
\bq
{H}_{2} := {H}_{1} + \RR_{1} {W}_{1}\ \ \ \ \ {V}_{2} := -
{e^{\{\Gamma_{1} {W}_{1}\}} - 1 \over \{\Gamma_{1}
{W}_{1}\}}.\Gamma_{1} \dot{W}_{1} \label{073}
\eq
and with \( \RR_{1}, \NN_{1}, \Gamma_{1}, {F}_{1} \) defined as
before, but with \( {H} \) replaced by \( {H}_{1} \). And so on.
Generally this iteration doesn't converge: so we must stop it at an
optimized order. This has been done in \cite{21}, with a different
method and framework.

Finally let us explicit the norm (\ref{052}) of the operator \( \Gamma
\) for the norm chosen in (\ref{051}):
\bq
|||\Gamma||| \leq\ \sup_{{A}, {A}'\ \st {A}\neq {A}'}\ {\psi({A},
{A}') \over |{h}({A}) - {h}({A}')|} \label{149}
\eq
where:
\bq
\psi({A}, {A}') := {g}_{{A}, {A}'}.\ \sup_{{A}''} \max \Biggl(
{{g}_{{A}, {A}''} \over {g}_{{A}', {A}''}}\ ,\ {{g}_{{A}', {A}''}
\over {g}_{{A}, {A}''}} \Biggr) \label{109}
\eq
and \( {g}_{{A}, {A}'} = {g}({A}, {A}' - {A}) \). The proof is an easy
estimation of the norm (\ref{052}). Hence the hypothesis 2 (\ref{053})
requires that:
\bq
\forall {A}, {A}'\ \ \st\ {A}\neq {A}'\ \ \ \ \ |{h}({A}) - {h}({A}')|
\geq \gamma. \psi({A}, {A}') \label{026}
\eq
for some constant \( \gamma > 0 \). This is a ``Diophantine''
condition. It depends on the choice of the weight function \( {g} \),
i.e. on the regularity of the perturbations \( {V} \) we want to
consider.

For instance, when \( \HH = \bbN^{*} \):
\bq
\If\ \ \ {g}_{{A},{A}'} = \varphi_{A} \cdot \varphi_{{A}'}\ \ \
\mathrm{then}\ \ \ \psi({A}, {A}') = \max (\varphi_{A}^{2},
\varphi_{{A}'}^{2}) \label{151}
\eq
As a particular case we can take, for some \( \alpha \geq 1 \):
\bq
{h}({A}) = {A}^{\alpha}\ \ \ \And\ \ \ \varphi_{A} = {A}^{\alpha - 1
\over 2} \label{152}
\eq
for which the condition (\ref{026}) is satisfied.

See \cite{14} for more details.

\section{- References}
\label{S05}


\vspace{-1.0cm}
\begin{thebibliography}{99}
\bibitem{07} J. BELLISSARD, M. VITTOT: ``Heisenberg's picture and non
commutative geometry of the semi classical limit in quantum
mechanics'', Annales de l'Institut Henri Poincar\'e, A, vol 52,
\underline{3}, (1990) p 175-235.
\bibitem{22} T. CARLETTI: ``The Lagrange inversion formula on
non-Archimedean fields. Non-Analytical Form of Differential and Finite
Difference Equations'', Discrete and Continuous Dynamical Systems,
Series A, Vol 9, 4, (July 2003), p 835-858. Archived in
http://arxiv.org/pdf/math/0110135
\bibitem{30} G. CIRAOLO, C. CHANDRE, R. LIMA, M. VITTOT, M. PETTINI:
``Control of chaos in Hamiltonian systems'', to appear in Cel. Mech.
\& Dyn. Astr. (2004). Archived in http://arxiv.org/nlin.CD/0311009
\bibitem{15} G. CIRAOLO, C. CHANDRE, R. LIMA, M. PETTINI, M. VITTOT,
Ch. FIGARELLA, Ph. GHENDRIH: ``Controlling chaotic transport in an
Hamiltonian model of interest to magnetized plasmas'',
J. Phys. A: Math. Gen. {37} (2004) 3589. Archived in
http://arxiv.org/pdf/nlin.CD/0304040 (2003).
\bibitem{25} G. CIRAOLO, F. BRIOLLE, C. CHANDRE, E. FLORIANI, R. LIMA,
M. VITTOT, M. PETTINI, C. FIGARELLA, P. GHENDRIH: ``Control of
Hamiltonian chaos as a possible tool to control anomalous transport in
fusion plasmas'', Phys. Rev. E {69} (4), (2004) 056213. Archived in
http://arxiv.org/pdf/nlin.CD/0312037
\bibitem{13} P. DUCLOS, O. LEV, P. STOVICEK, M. VITTOT: ``Progressive
diagonalization and applications'', Proceedings of the Conference
``Operator Algebras \& Mathematical Physics'', Constan\c ta (Roumanie,
2001), R. Purice Ed., Theta Foundation, Bucarest (2003). Archived in
http://arxiv.org/pdf/math-ph/0405014
\bibitem{14} P. DUCLOS, O. LEV, P. STOVICEK, M. VITTOT: ``Weakly
regular Floquet Hamiltonians with pure point spectrum'',
Reviews in Mathematical Physics, \underline{14} 6 (2002), p 1-38.
Archived in http://arxiv.org/pdf/math-ph/0103041
\bibitem{23} L.H. ELIASSON: ``Hamiltonian systems with linear normal
form near an invariant torus'', in ``Non-Linear Dynamics'' (Bologne,
1988), G. Turchetti Ed., World Scientific, Singapore (1989), p 11-29.
\bibitem{16} R.S. HAMILTON: ``The inverse function theorem of Nash and
Moser'', Bull. Amer. Math. Soc. (N.S.), 7 (1), (1982), p 65-222.
\bibitem{21} A. JOYE, Ch.-Ed. PFISTER: ``Exponential Estimates in
Adiabatic Quantum Evolution'', International Congress of Mathematical
Physics (ICMP 1997), De Wit, Bracken, Gould \& Pearce Eds,
International Press, Cambridge. 6 (1999), p.309-315
\bibitem{18} J. MOSER: ``A new technique for the construction of
solutions of nonlinear differential equations'', Proc. Nat. Acad. Sci.
U.S.A., 47, (1961), p 1824-1831.
\bibitem{17} J. NASH: ``The imbedding problem for Riemannian
manifolds'', Ann. of Math. (2), 63, (1956), p 20-63.
\bibitem{19} M. PETTINI: ``Low Dimensional Hamiltonian Models for
Non-Collisional Diffusion of Charged Particles'', in ``Non-Linear
Dynamics'' (Bologne, 1988), G. Turchetti Ed., World Scientific,
Singapore (1989), p 287-296.
\bibitem{08} M. VITTOT: ``Lindstedt perturbation series in Hamiltonian
mechanics: explicit\\ formulation via a multidimensional
B\"urmann-Lagrange formula'', Preprint CNRS Luminy CPT-91/P.2603,
Marseille (1991). Archived in\\
http://ccdb3fs.kek.jp/cgi-bin/img/allpdf?199204265
\end{thebibliography}
\end{document}